\documentclass{article}
\usepackage{graphicx}
\usepackage{url}

\newcommand{\figlab}[1]{\label{fig:#1}}

\newcommand{\figref}[1]{\ref{fig:#1}}
\newcommand{\eqref}[1]{(\ref{eq:#1})}

{\catcode`\@=11
\gdef\setft#1#2#3{%
\def\@oddfoot{
{\setbox0=\hbox{#1}
\setbox1=\hbox{#3}
\ifdim\wd0>\wd1
\dimen0=\wd0
\box0\hfil#2\hfil\hbox to\dimen0{\hfil\hfil\box1}
\else \dimen0=\wd1
\hbox to\dimen0{\box0\hfil }\hfil#2\hfil\box1 \fi
}}} }

\def\complaint#1{}
\def\withcomplaints{
\newcounter{mycomplaints}
\def\complaint##1{\refstepcounter{mycomplaints}%
\ifhmode%
\unskip%
{\dimen1=\baselineskip \divide\dimen1 by 2 %
\raise\dimen1\llap{\tiny -\themycomplaints-}}\fi%
\marginpar{\tiny [\themycomplaints]: ##1}}%
}

\usepackage{amssymb}

\title{Computational Geometry Column 39}
\author{%
Joseph O'Rourke\thanks{
Dept. of Computer Science, Smith Col\-lege, North\-ampton, 
MA 01063, USA.
\-orourke@cs.\-smith\-.edu.
Supported by NSF Grant CCR-9731804.
}
}
\date{}
\begin{document}
\maketitle
\pagestyle{empty}
\thispagestyle{empty}

\begin{abstract}
The resolution of a decades-old open problem 
is described:  polygonal chains cannot lock in the plane.
\end{abstract}

A {\em polygonal chain\/} is a connected series of line segments.
Chains may be open, or closed to form a polygon.
A {\em simple\/} chain is one that does not self-intersect:
only segments adjacent in the chain intersect, and then
only at their shared endpoint.
If the segments of a polygonal chain are viewed as
rigid bars, and the vertices as universal joints, natural
questions are whether every open chain can be
{\em straightened\/}---reconfigured to lie on a straight line---and 
whether every closed chain can be {\em convexified\/}---reconfigured
to form a planar convex polygon.  In both cases, the chains are
to remain simple throughout the motion.
If a chain cannot be so reconfigured, it is called {\em locked}.

These questions were raised
by several researchers independently since the 1970's,%
\footnote{
	G.~Bergman,
	U.~Grenander,
	W.~Lenhart,
	J.~Mitchell,
	S.~Schanuel, 
	and 
	S.~Whitesides.
	It seems to have first appeared in print in 1995:
	\cite{lw-rcpce-95} and~\cite[p.~270]{k-pldt-95}.
}
and were the subject of intense investigation by the late 1990's.
It was first established that chains can lock in three dimensions (3D):
both locked open chains, and locked closed but unknotted chains
are possible~\cite{cj-nepiu-98,bddlloorstw-lupc3d-99}.
In 4D, neither open nor closed chains can 
lock~\cite{co-pccl4d-99}.
But aside from some special cases 
(e.g., star-shaped polygons cannot lock~\cite{elrss-cssp-98},
polygonal trees can lock~\cite{bddloorsw-orfltcl-98}),
the problems remained unresolved for chains in 2D.

Connelly, Demaine, and Rote have now settled the questions,
establishing that neither open nor closed chains can lock
in 2D~\cite{cdr-epcbu-00}.  
Their result is even more general:  no collection
of disjoint simple chains are locked
(although of course a chain nested inside a polygon is forever
confined).
They prove that every such collection has an
{\em expansive motion\/}: one during which the distance
between every pair of vertices increases or stays the same.
An expansive motion automatically maintains simplicity,
for the distance between some pair of vertices would have to decrease
to reach self-intersection.
Their proof first establishes that any configuration of
chains has an infinitesimal expansive motion,
and then combines these motions into a global expansive motion.
Although there are different ways to stitch the infinitesimal
motions together, one produces especially
natural movements.
This motion 
minimizes the squared lengths
of the velocity vectors $v_i$ applied to each vertex $i$ at each time:
minimize $\sum_i \| v_i \|^2$, subject to constraints maintaining
segment lengths and forcing expansivity.\footnote{
	Their proof employs a further nonquadratic 
	``barrier'' penalty term
	for smoothness.
}
The objective function is strictly convex and the system
can be solved by quadratic programming.
Figs.~\figref{chains.1} and~\figref{chains.2} 
illustrate the solution so obtained for
a collection of three closed chains (triangles) entangled with one open
chain.
\begin{figure}[htbp]
\centering
\includegraphics[width=0.2\linewidth]{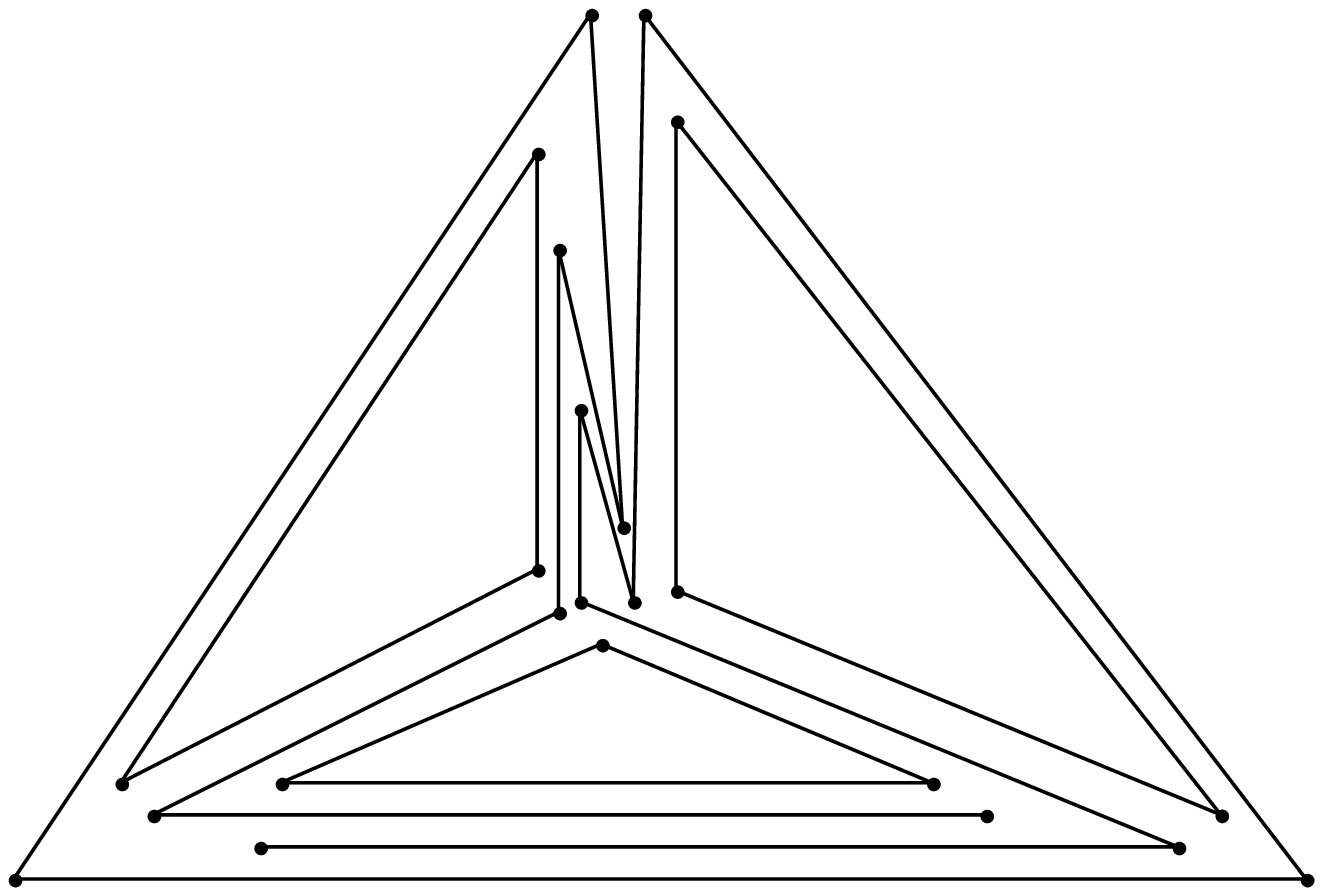}%
\hspace{2mm}%
\includegraphics[width=0.2\linewidth]{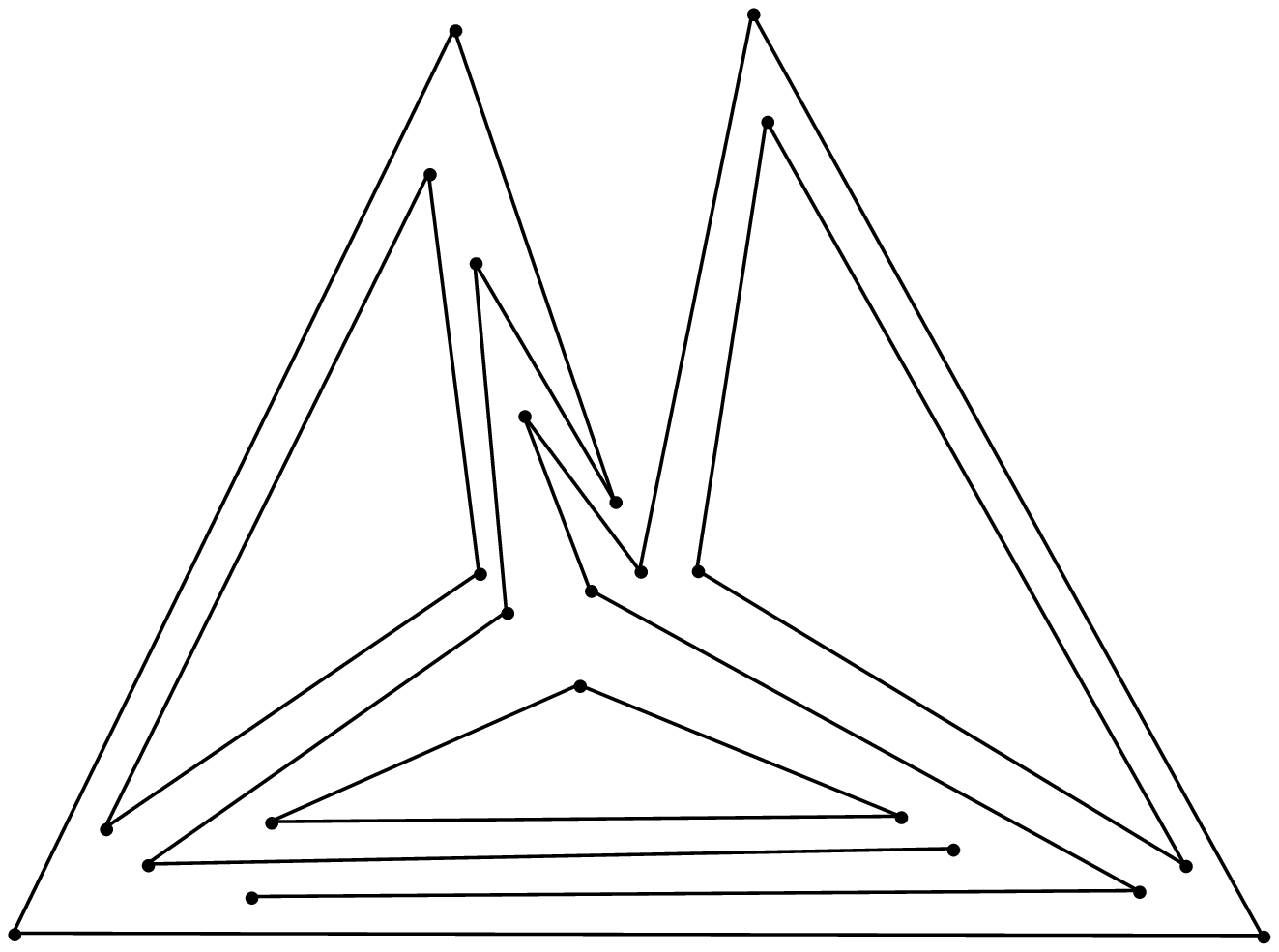}%
\hspace{2mm}%
\includegraphics[width=0.2\linewidth]{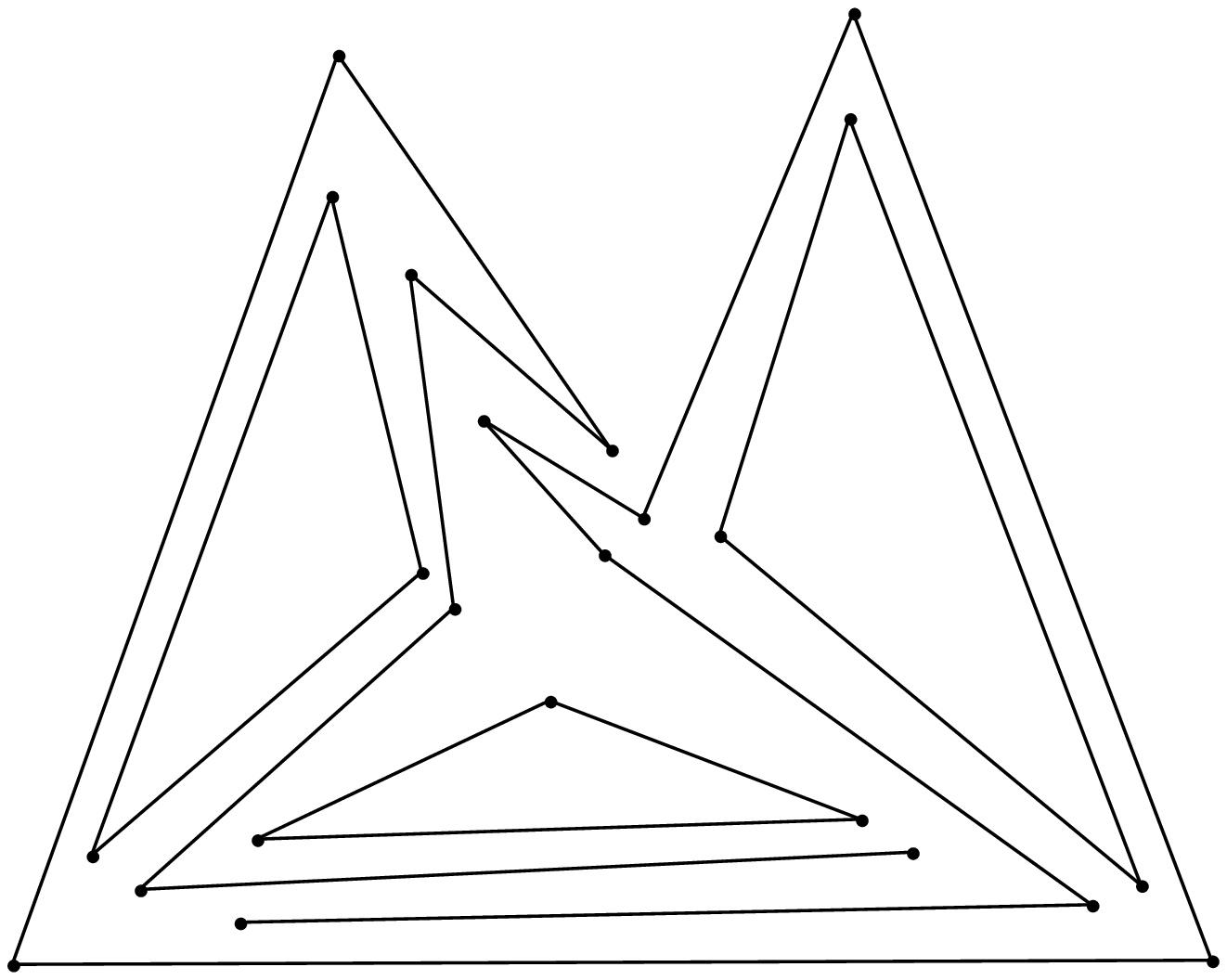}%
\hspace{2mm}%
\includegraphics[width=0.2\linewidth]{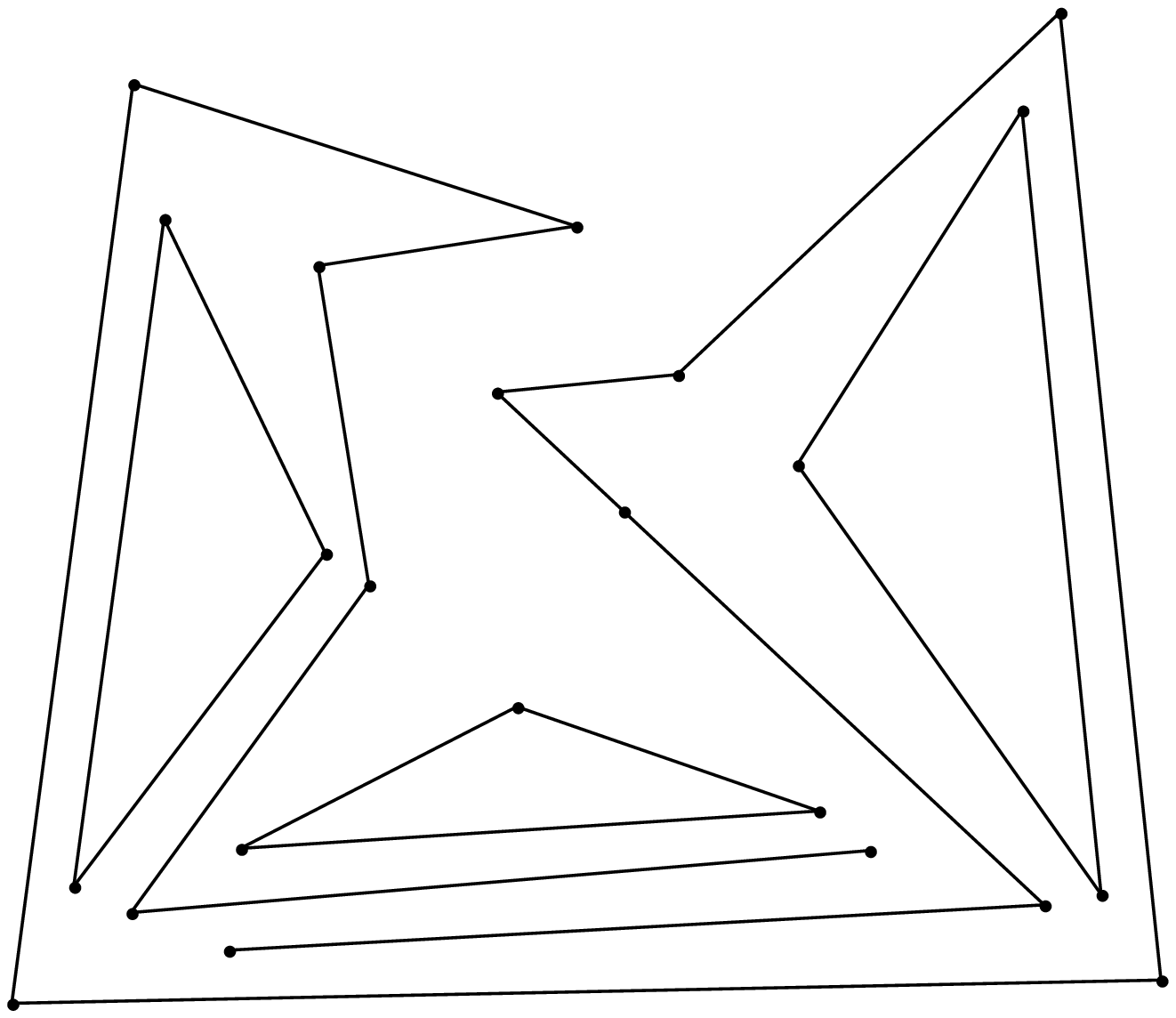}
\vspace{-2mm}
\caption{Four frames in the unlocking of four polygonal chains.
The initial disjoint collection is shown in the leftmost frame.
[Animation courtesy of Erik Demaine.]}
\figlab{chains.1}
\centering
\includegraphics[width=0.25\linewidth]{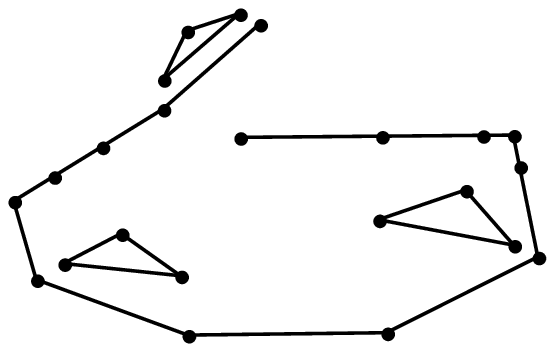}%
\hspace{1mm}%
\includegraphics[width=0.25\linewidth]{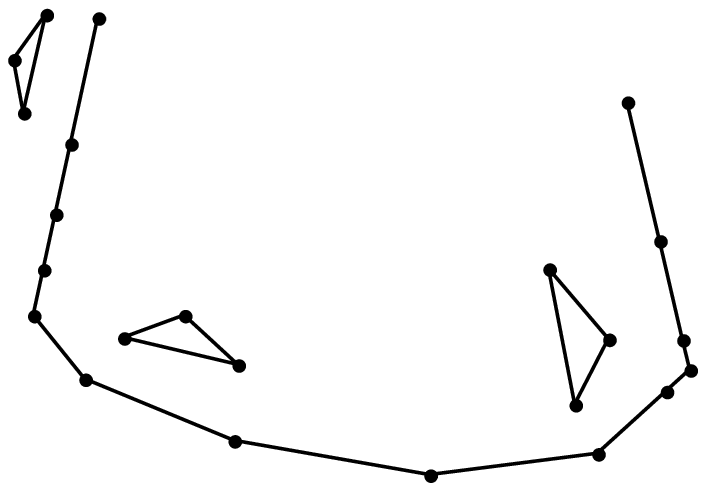}%
\hspace{5mm}%
\includegraphics[width=0.25\linewidth]{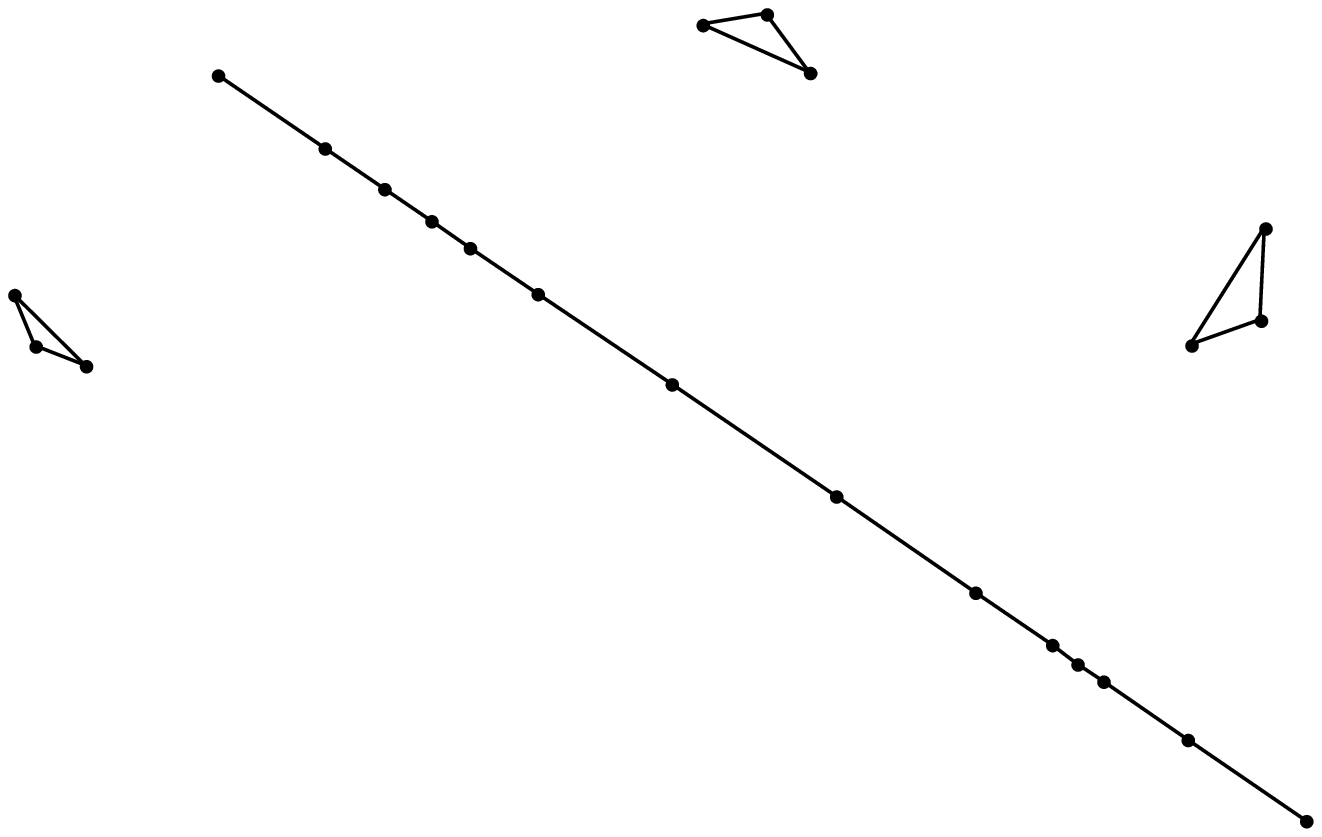}%
\vspace{-8mm}
\caption{Three frames toward the end of the
unlocking motion, continuing Fig.~\figref{chains.1}
(at a different scale).}
\figlab{chains.2}
\end{figure}

Let us simplify the discussion to a single open chain
and offer a crude sketch of their key argument.
View nonadjacent vertices of the chain as connected by
a {\em strut}, which is permitted to increase in length
or stay the same, but never decrease.
If this bar-and-strut framework has an infinitesimal
motion, then the motion is necessarily expansive.
An {\em equilibrium stress\/} in an assignment of
weights ({\em stresses\/}) to each bar and strut so that (a)~the stress on
each strut is nonnegative, and (b)~the stress vectors 
(vectors along the bars/struts scaled by the stresses) are
in equilibrium at each vertex.  
Bar stresses may be positive (compression)
or negative (tension).
If a framework 
is ``stuck,'' or, more formally, 
``first-order rigid''~\cite{cw-sorps-96,w-rsa-97},
then it has a nonzero equilibrium stress.
So if a framework only has the trivial zero
equilibrium stress, then it possesses an
infinitesimal expansive motion.

In order to apply a theorem that holds for planar frameworks,
every intersection point of the framework is used to
divide the bars/struts, which produces a planar
framework 
equivalent to the original in
terms of equilibrium stresses.
Now the century-old Maxwell-Cremona theorem is applied:
if the framework has a nonzero equilibrium stress,
then it can be ``lifted'' to a nonflat polyhedral terrain
that projects to the framework, with positive-stress edges
(always bars)
lifting to ``mountains,'' and negative-stress edges to ``valleys.''
Finally it is shown that such a lifting is impossible,
by concentrating on its maximum.
For example, if this maximum is a single point,
then it must lie at the junction of at least three
mountain-edges, which violates the fact that
every vertex of the chain is incident to at most two bars.
A case analysis on the topologically possible maxima
proves that only the degenerate flat lifting
is possible, which implies that the original framework
has only the zero 
equilibrium stress, which implies that it has an 
infinitesimal expansive motion,
which implies that it has a global expansive motion,
i.e., it is not locked.

Building on this work, Streinu has found a way to
decompose the expansive global motion for chains of $n$ vertices
into $O(n^2)$ sections, each of which is the motion of
a one degree-of-freedom mechanism~\cite{s-capnc-00}.
This mechanism is constructed by removing one hull edge
from a {\em pseudotriangulation\/} of the chains, 
a partition by diagonal bars into regions each bounded by
three reflex chains (where a single segment counts as a reflex chain).
The mechanism is opened following its only free trajectory
in configuration space
until two adjacent edges align,
at which point the pseudotrianguation is revised locally, and
the expansion process continued.

\bibliographystyle{alpha}
\bibliography{39}
\end{document}